\documentclass[aps,prl,superscriptaddress,twocolumn,10pt]{revtex4-1}

\usepackage{ucs}
\usepackage[latin1]{inputenc}
\usepackage[english]{babel}
\usepackage[usenames, dvipsnames]{xcolor}
\usepackage{amsmath, 
			amsfonts,
            booktabs,
            microtype,
            mathtools, 
            graphicx,
            color,
            nicefrac,
            url, 
            placeins,	%float barrier
			braket,
			appendix} 	%chemie kram

\DeclareMathOperator{\sign}{sign}

\begin{document}
\title{Two-body interactions and decay of three-level Rydberg-dressed atoms}
\author{S. Helmrich}\affiliation{Physikalisches Institut, Universit\"at Heidelberg, Im Neuenheimer Feld 226, 69120 Heidelberg, Germany.}
\author{A. Arias}\affiliation{Physikalisches Institut, Universit\"at Heidelberg, Im Neuenheimer Feld 226, 69120 Heidelberg, Germany.}
\author{N. Pehoviak}\affiliation{Physikalisches Institut, Universit\"at Heidelberg, Im Neuenheimer Feld 226, 69120 Heidelberg, Germany.}
\author{S. Whitlock}\email{whitlock@uni-heidelberg.de}\affiliation{Physikalisches Institut, Universit\"at Heidelberg, Im Neuenheimer Feld 226, 69120 Heidelberg, Germany.}
\date{\today}
	
\begin{abstract}
	We theoretically analyze the interactions and decay rates for atoms dressed by multiple laser fields to strongly interacting Rydberg states using a quantum master equation approach. In this framework a comparison of two-level and three-level Rydberg-dressing schemes is presented. We identify a resonant enhancement of the three-level dressed interaction strength which originates from cooperative multiphoton couplings as well as small distance dependent decay rates. In this regime the soft-core shape of the potential is independent of the sign of the bare Rydberg-Rydberg interaction, while its sign can be repulsive or attractive depending on the intermediate state detuning. As a consequence, near-resonant Rydberg dressing in three-level atomic systems may enable the realization of laser driven quantum fluids with long-range and anisotropic interactions and with controllable dissipation. %beyond what can be achieved via two-level dressing.
\end{abstract}
	
\maketitle

% Introduction 
%%%%%%%%%%%%%%%%%%%%%%%%%%%%%%%%%%%%%%%%%%%%%%%%%%%%%%%%%%%%%%%%%%%%%%%%%
Ultracold atomic gases are versatile model systems for investigating novel phases of matter and condensed-matter phenomena with nearly complete control over the relevant system parameters. To date however, the landmark experiments in the field, i.e. the realization of Bose-Einstein condensation \cite{Anderson1995,Davis1995,Bradley1995} and degenerate Fermi gases with molecular or Bardeen-Cooper-Schrieffer correlations~\cite{Greiner2003,Jochim2003,Zwierlein2003,Regal2004,Zwierlein2004,Kinast2004,Bourdel2004,Chin2004} are governed by isotropic contact interactions. This is an important distinction from, for example, the dominant Coulomb interactions in strongly-correlated electron systems~\cite{Dagotto2005} or the complex interactions which govern nematic liquid crystal phases~\cite{Gennes1993}. This apparent limitation is fading though with the advent of ultracold gases of atoms possessing sizeable magnetic dipole moments~\cite{Lahaye2009,Stuhler2005,Aikawa2012,Lu2012} or heteronuclear molecules with electric dipole moments~\cite{Deiglmayr2010,Ni2008,Takekoshi2014}. In these systems the long-range and anisotropic character of the interactions gives rise to intriguing effects such as dipolar collapse \cite{Lahaye2008}, demagnetization dynamics~\cite{Sadler2006,Pasquiou2011}, universal dipolar scattering \cite{Aikawa2014b}, roton instabilities \cite{Kadau2015} as well as magnetic or Hubbard-like physics arising from beyond contact interactions \cite{DePaz2013,Baier2015,Yan2013}. Future experiments aim to further enhance and control these interactions, opening the door to new types of strongly-correlated dipolar matter such as topological phases~\cite{Yao2012,Li2015,Peter2014,VanBijnen2014,Glaetzle2015} or states simultaneously exhibiting crystalline and superfluid order~\cite{Henkel2010,Pupillo2010,He2011,Geissler2015}. 

Highly excited Rydberg states of atoms offer another promising approach to create strongly-correlated matter by engineering the strength, range and anisotropy of interparticle interactions. However, the energy scales associated with the Rydberg-Rydberg interactions (tens of MHz) and their relatively short lifetimes ($\sim 10~\mu$s) are seemingly incompatible with the typical energy scales ($\sim$~kHz) and time scales associated with ultracold atomic motion or equilibration ($\sim 10~\mathrm{ms}$). A possible resolution, first proposed by Santos et. al.~\cite{Santos2000}, is to treat each particle as composed of the atomic ground $\ket{g}$ and Rydberg $\ket{r}$ states coherently coupled by a far off-resonant laser. The resulting \emph{dressed} atoms $\ket{\psi}\approx \ket{g}+\beta\ket{r}$ include a small Rydberg-state admixture ($\beta\ll 1$) which would allow them to evolve under the influence of strong and long-range interactions for up to tenths of a second. So far, theoretical and experimental studies have employed a two-photon excitation scheme with a large detuning from an intermediate state (effective two-level regime)~\cite{Henkel2010,Pupillo2010,Honer2010,Johnson2010,Maucher2011,Hamadeh2012,Henkel2012,Balewski2014,Geissler2015,Carmele2014} or direct excitation via a single laser field~\cite{Jau2015,Wang2014,Wuster2011a,Mobius2013,Macri2014,Glaetzle2015}. However, experimentally reaching conditions where coherence times exceed motional timescales still remains an important challenge. 

\begin{figure*}[!t]
	\centering
	\includegraphics[width = 1.65\columnwidth]{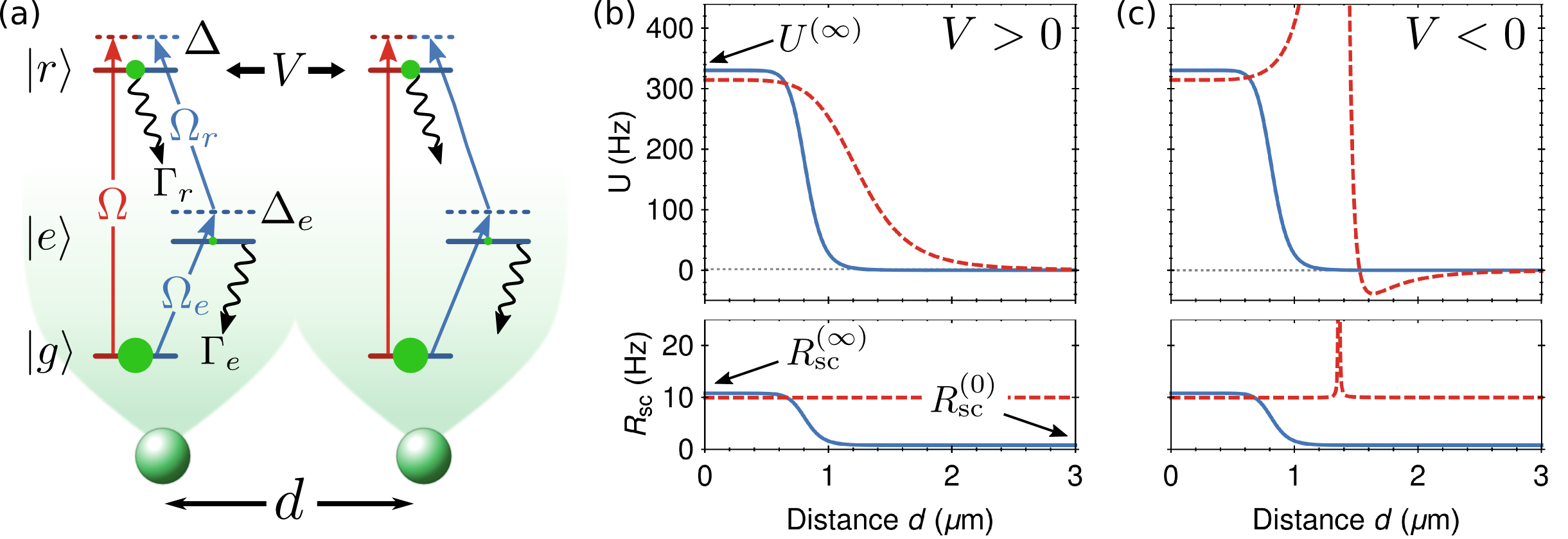}
	\caption{(a) Level scheme for two atoms with distance $d$ dressed by one or two laser fields colored red and blue respectively. (b) Comparison of the interaction strength $U(C_6/d^6)$ and scattering rate $R_\text{sc}(C_6/d^6)$ for two-level (dashed red lines) and three-level systems (solid blue lines) assuming repulsive van der Waals interactions between Rydberg states and $\Delta=0,\Delta_e>0$ (three-level) and $\Delta<0$ (two-level). (c) Same as in (b) but for attractive van der Waals interactions. The parameters used for the calculations are given in the text.}
	\label{fig:2-phot_dressing-curves}
\end{figure*}
Here we treat Rydberg dressing including the larger parameter space afforded by multiple dressing laser fields. Furthermore we present a detailed comparison of two-level and three-level Rydberg-dressing schemes including the influence of spontaneous decay of the excited states. For a wide range of parameters the effective three-level interaction potentials have a soft-core form similar to the two-level case. The optimal parameters for three-level dressing coincide with destructive interference of the intermediate state population and a cooperative enhancement of the Rydberg admixture due to multiphoton excitations \cite{gaerttner2014}. For these parameters the dressed-state interaction strength and lifetime can be comparable to, or even exceed what is possible for two levels. Remarkably the shape of the dressed potential is independent of the sign of the bare Rydberg-Rydberg interaction, opening the possibility to realize stable Rydberg dressed atoms with anisotropic interactions.  

To self-consistently calculate the effective interaction potential $U$ and the residual photon scattering rate $R_\text{sc}$ as a function of the bare Rydberg-Rydberg interaction strength $V$ we use a quantum master equation treatment which includes spontaneous decay from the excited states. The master equation describing the two-particle system is 
$\dot \rho=-i [\hat H, \rho]+\mathcal{L}[\rho]$ (in units where $\hbar=1$). The Hamiltonian consists of three parts  $\hat H=\hat H_1 + \hat H_2 + \hat V$, where $\hat H_{1,2}$ are the individual atom-light Hamiltonians in the rotating wave approximation and $\hat V = V \ket{r_1}\ket{r_2}\bra{r_1}\bra{r_2}$ is the two-body interaction between bare Rydberg states. The superoperator $\mathcal{L}[\rho]=\sum_{\{ \hat L \}} \hat L\rho \hat L^\dagger-(\hat L^\dagger\hat L\rho+\rho\hat L^\dagger\hat L)/2$ describes spontaneous decay of the excited states, where $\{ \hat L \}$ is a set of decay operators. Focusing on the three-level ladder system (Fig~\ref{fig:2-phot_dressing-curves}a), in which each atom is composed of a long lived ground state $\ket{g}$ coupled to the Rydberg state $\ket{r}$ via a short lived intermediate state $\ket{e}$, then
%%%%%%%%%%
\begin{eqnarray} \hat H_j&=&\frac{\Omega_e}{2}\ket{g_j}\bra{e_j}+\frac{\Omega_r}{2}\ket{e_j}\bra{r_j}\nonumber \\
& &-\frac{\Delta_e}{2}\ket{e_j}\bra{e_j}-\frac{\Delta}{2}\ket{r_j}\bra{r_j}+\text{H.c.}
\end{eqnarray}
%%%%%%%%%%
Additionally,  $\hat L_{j}=\{\sqrt{\Gamma_e}\ket{g_j}\bra{e_j}, \sqrt{\Gamma_r}\ket{e_j}\bra{r_j}\}$ describes spontaneous decay from the intermediate state and the Rydberg state respectively. We neglect possible couplings to other Rydberg states and the much weaker interactions between atoms in the $\ket{g}$ or $\ket{e}$ states. Typically $\Gamma_e \gg \Gamma_r$ and the internal degrees of freedom reach steady state on a timescale $\sim \Gamma_e^{-1}$ which we assume is much faster than typical motional timescales. Therefore it is sufficient to calculate the steady state values of the dressed potential and residual scattering rate for each value of $V$ without explicitly considering the motion of the particles. However care must be taken applying the same reasoning to the two-level case or to atoms with metastable intermediate states for which internal state dynamics can be much slower. To calculate $U(V)$ we first solve the master equation for steady state $(\dot\rho=0)$ and then compute the expectation value $U(V) = \text{Tr}[\hat H\rho]-\text{Tr}[\hat H\rho]_{V=0}$, where the last term subtracts the single particle light shifts. Similarly, the dressed-atom decay rate (per atom) responsible for heating and loss of dressed atoms is given by $R_\text{sc}=(1/2)\sum_{\{ \hat L \} } \text{Tr}[\hat L \rho \hat L^\dagger]$.

Generally, the inclusion of the intermediate state can have a dramatic influence on the shape and strength of the dressed interactions as well as the photon scattering rate as a function of the interparticle separation $d$. Figures~\ref{fig:2-phot_dressing-curves}b,c show calculated two-body dressed state potential energy surfaces $U(C_6/d^6)$ and decay rates $R_\mathrm{sc}(C_6/d^6)$ for the two-level and three-level systems assuming van der Waals interactions with strength $C_6$ and the following achievable experimental values: $C_6/2\pi=\pm 1.0~$GHz~$\mu$m$^6$, $\Gamma_e/2\pi=6.0$~MHz, $\Gamma_r/2\pi=10$~kHz,  $\Omega_e/2\pi = 1.2~\text{MHz}$, $\Omega_r/2\pi = 200~\text{MHz}$, $\Delta_e/2\pi = 100~\text{MHz}$, $\Delta = 0$ (three-levels) and $\Omega/2\pi = 5.0~\text{MHz}$, $\Delta/2\pi = -79~\text{MHz}$ (two-levels). These parameters were chosen such that both systems exhibit similarly small decay rates ($\lesssim 10$~Hz) and long-range effective interactions which `soften' to a constant value $U^{(\infty)}$ at short distances as a consequence of the Rydberg blockade~\cite{Lukin2001}. We note however that the three-level dressed potential is considerably more box-like than for two-levels, and its range is shorter due to the larger excitation bandwidth resulting in a smaller blockade radius. 

\begin{figure}[t!]
	\centering
	\includegraphics[width = 1.0\columnwidth]{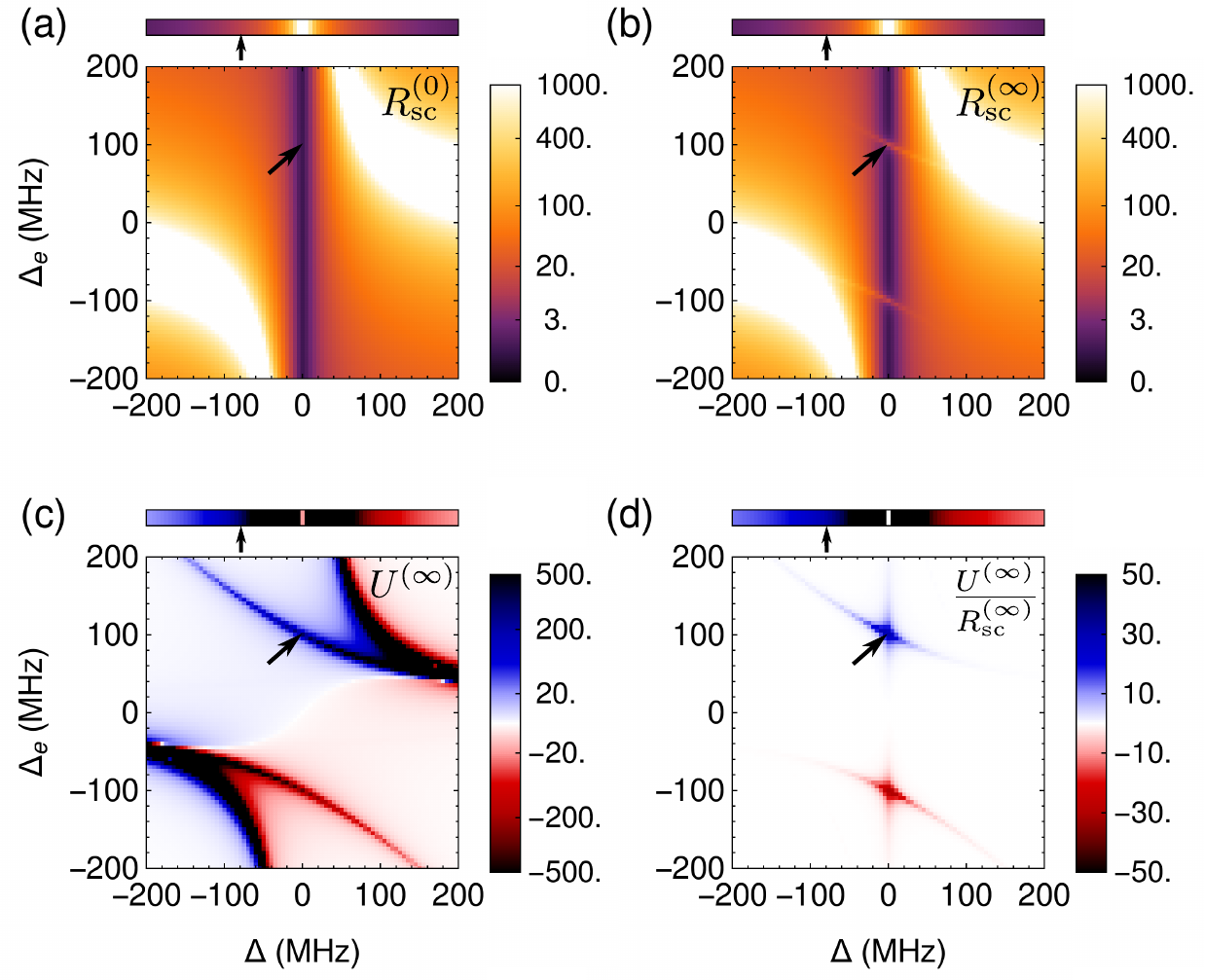}
	\caption{Dressed state decay rates and interaction strength as a function of the laser detunings. (a) Decay rate per atom $R_\text{sc}^{(0)}$ for $V\rightarrow 0$ corresponding to two independent particles. (b) Decay rate per atom $R_\text{sc}^{(\infty)}$ for $V\rightarrow \infty$ corresponding to the fully blockaded limit. (c) Dressed state interaction strength $U^{(\infty)}$. Blue corresponds to repulsive and red to attractive interactions. (d) Figure of merit $f=U^{(\infty)}/R_\text{sc}^{(\infty)}$. The colorscales for figures (a-c) follow a power-law to emphasize small features whereas (d) is on a linear scale. The horizontal bars above each plot show the corresponding two-level parameters on the same colorscale as a function of the detuning $\Delta$. The arrows indicate the detunings used in Fig.~\ref{fig:2-phot_dressing-curves} and are described in the text.}
	\label{fig:2-phot_dressing-2Dplots}
\end{figure}	
A distinguishing feature of three-level dressing is that the decay rate can be strongly spatially dependent, which is in contrast to the two-level case which only shows a small reduction of $R_\text{sc}$ at short distances. We focus on two limits: $V=0$ and $V\rightarrow \infty$. For large interparticle separations or dilute-gas experiments the residual decay rate $R_\text{sc}^{(0)}=R_\text{sc}(V=0)$ is minimized due to destructive interference of the intermediate state population due to electromagnetically induced transparency (EIT). For short distances or in dense gases on the other hand the Rydberg blockade effect breaks the EIT condition leading to increased decay rates which plateau at a value $R_\text{sc}^{(\infty)}=R_\text{sc}(V\rightarrow \infty)$. This enhanced distance dependent decay may find applications in quantum state engineering via dissipation \cite{Schempp2015} or novel cooling techniques \cite{Glaetzle2012,Zhao2012,Huber2012}. Another important difference between the three-level and two-level cases concerns the dependence on the sign of the bare-state interactions. In the two-level case the sign of $V$ and $\Delta$ must be opposite to avoid level crossings which deform the potential and give rise to strongly enhanced decay (Fig.~\ref{fig:2-phot_dressing-curves}c). In contrast, the shape and sign of the potential in the three-level case can be made independent of the sign of $V$ and can be manipulated by $\Delta_e$.  
	
While the two-level case has relatively few tuning parameters ($\Omega$ and $\Delta$), the three-level case presents additional possibilities through independently tuning $\Omega_e,\Omega_r,\Delta_e$ and $\Delta$. Therefore we search for optimal parameters which maximize $U^{(\infty)}$ while keeping $R_\mathrm{sc}$ small. To ensure a small Rydberg state population we exploit Autler-Townes splitting, focusing on the parameter regime $\Omega_{r}\gg\Gamma_e \gg \Gamma_r$ and $\Omega_r \gg \Omega_e$. The qualitative features of $R_\text{sc}$ and $U^{(\infty)}$ in this parameter regime are mostly independent of the Rabi frequencies which predominantly influence the overall energy and time scales. Figures~\ref{fig:2-phot_dressing-2Dplots}a,b show the characteristic decay rates $R_\mathrm{sc}^{(0)}$, $R_\mathrm{sc}^{(\infty)}$ for the three-level system as a function of the intermediate- and two-photon detunings $\Delta_e$ and $\Delta$ respectively. The Autler-Townes doublet is clearly seen as bright bands for $\Delta_e=\Omega_r^2/(4\Delta)$ and the distinct minimum for $\Delta=0$ is due to destructive interference of the intermediate state amplitude. In the interaction dominated regime we observe two additional features in $R_\text{sc}^{(\infty)}$ for $\Delta_e=-(\Delta\pm\sqrt{\Delta^2+\Omega_r^2})/2$ (Fig.~\ref{fig:2-phot_dressing-2Dplots}b). These features correspond to cooperative resonances between the two-atom ground state and the $\ket{e_1}\ket{e_2}$, $\ket{e_1}\ket{r_2}$ and $\ket{r_1}\ket{e_2}$ states. 

Figure~\ref{fig:2-phot_dressing-2Dplots}c shows the interaction strength $U^{(\infty)}$ which is split into two domains of repulsive (blue) and attractive (red) interactions. Maximum interaction strengths are found on the Autler-Townes resonances (coinciding with maximal decay rates) and on the cooperative resonances (with small decay rates). This cooperative enhancement of the dressed-state interactions does not appear in an effective two-level description. A similar type of enhancement exploiting molecular resonances for specific pair distances and Rydberg states has recently been proposed~\cite{VanBijnen2014}, however the cooperative enhancement reported here works for any Rydberg state and preserves the soft-core nature of the dressed potential. We also expect it to persist for more than two atoms~\cite{gaerttner2014}, therefore it may also prove beneficial for the implementation of collective many-body interactions in Rydberg dressing~\cite{Balewski2014}. Figure~\ref{fig:2-phot_dressing-2Dplots}d shows our chosen figure-of-merit for Rydberg dressing $f = U^{(\infty)}/R_\mathrm{sc}^{(\infty)}$ as a function of the two detunings. The optimal detunings are clearly revealed as dark crosses at the intersection of the two-photon resonance line and the cooperative resonances at $\left(\Delta,\Delta_e) = (0, \pm \Omega_r/2\right)$ which coincides with the parameters chosen for Fig.~\ref{fig:2-phot_dressing-curves}b,c. For the depicted parameters $f\approx 30$. For gases in the dilute regime we expect $R_\mathrm{sc}^{(0)}$ to be the relevant decay rate, in which case the associated figure of merit can even be a factor of 10 or more larger. 

\begin{table}[t!]
\centering
\begin{ruledtabular}
\begin{tabular}{lcc}
						& {\bf 2-level} & {\bf optimized 3-level} \\
	\hline\vspace{-2mm} \\
	$\beta^2$ 			& $\Omega^2/(2\Delta)^2$ & $\Omega_e^2/\Omega_r^2$ \\ 
	$R_\text{sc}^{(0)}$ & $\Gamma_r \beta^2$ & $2 \Gamma_r \beta^2$ \vspace{1mm}\\ 
	 $R_\text{sc}^{(\infty)}$ 	& $-2\Gamma_r \beta^4 + R_\mathrm{sc}^{(0)}$ & \color{black}$\beta^4\Omega_r^2/\Gamma_e + R_\mathrm{sc}^{(0)}$  \vspace{1mm}\\ 
	$U^{(\infty)}$ 		& $-2\beta^3 \Omega$ & \color{black} $\sign (\Delta_e) \beta^4 \Omega_r^3/{\Gamma_e}^2$ \vspace{1mm}\\
	$|f|$					& $\frac{2\Omega}{\Gamma_r} \left[ \beta^{-1}-2\beta \right]^{-1}$ & \color{black} $\frac{\Omega_r}{\Gamma_e} \left[ 1+\frac{2 \Gamma_r \Gamma_e}{\beta^2 \Omega_r^2} \right]^{-1}$ \vspace{1mm} \\
\end{tabular}
\end{ruledtabular}
\caption{Key parameters and approximate scaling relations characterising the dressed-state potential in the weak dressing regime: Rydberg fraction $\beta^2$, residual decay rates $R_\text{sc}^{(\infty)}$ and $R_\text{sc}^{(0)}$, interaction potential $U^{(\infty)}$ and the figure of merit $f = U^{(\infty)}/R_\text{sc}^{(\infty)}$. We restrict our analysis to $\beta\ll 1$ and $\sign(V)\neq \sign(\Delta)$ (two-level case) and $\Delta=0$ and $\Delta_e=\pm\Omega_r/2$ (optimized three-level case).
}
\label{tab:scalings}
\end{table}

We now discuss the scalings of the interaction strength and decay rate for the detunings corresponding to the two-photon resonance as well as the cooperative resonance and as a function of the remaining system parameters as summarised in Table~\ref{tab:scalings}. In the non-interacting limit the Rydberg fraction on two-photon resonance is $\beta^2 = \sum_{j}\mathrm{Tr} \left[\ket{r_j} \bra{r_j} \rho\right]/2 \approx \Omega_e^2 / \Omega_r^2$ as expected for the EIT dark state~\cite{Fleischhauer2005}. Remarkably, the decay rate is independent of $\Gamma_e$ and is given by $R_\text{sc}^{(0)} \approx 2 \Gamma_r \beta^2$. Both the interaction strength and the decay rate are maximal at the cooperative resonance with $\big|U_\mathrm{opt}^{(\infty)} \big| \approx \beta^4 \Omega_r^3/ \Gamma_e^2$ and $R_\mathrm{sc}^{(\infty)}\approx \beta^4\Omega_r^2/\Gamma_e+R_\text{sc}^{(0)}$, which for small $\Gamma_r$ is dominated by the first term. Comparing these two yields a figure of merit $|f|\approx \Omega_r/\Gamma_e$. This shows that the optimal condition for three-level dressing requires small intermediate-state decay rates and large couplings on the upper transition, but is independent of the Rydberg state admixture and the dressed state lifetime, which is free to be chosen through the ratio $\Omega_e^2/\Omega_r^2$. Corresponding expressions for the two-level system are provided in Table~\ref{tab:scalings}. By comparing the scalings of two- and three-level dressing for equal decay rates we conclude that the figure of merit for three-level dressing may outperform two-level dressing if 
\begin{equation}
\frac{\Omega_r}{\Gamma_e} > \frac{2\Omega}{\Gamma_r}\left(\frac{R_\text{sc}}{\Gamma_r}\right)^\frac{1}{2}.
\end{equation}

In conclusion, we have analyzed effective two-body interaction potentials and decay rates for Rydberg-dressed atoms coupled by multiple laser fields using a master-equation treatment. Although we primarily focused on two-level and three-level dressing schemes, this approach is also applicable to systems with more levels and more than two coupling fields. While two-level dressing (i.e. employing a single laser field coupling ground and Rydberg states) completely eliminates the population of short-lived states, which may be important for achieving long dressed-state lifetimes, we also find favorable conditions in the three-level system by making use of EIT interference on two-photon resonance. Additionally, we have identified optimum laser parameters corresponding to a cooperative enhancement of the dressed state potential which is not present for the effective two-level descriptions considered previously. The figure of merit for three-level dressing under these conditions scales with $\Omega_r/\Gamma_e$, indicating that large coupling strengths between intermediate and Rydberg states and long lifetimes of the intermediate state are desirable. This highlights the importance of high power lasers and large matrix elements for the upper transition, combined with long intermediate state lifetimes, such as is possible using the $\ket{ns}\rightarrow \ket{(n+1)p} \rightarrow \ket{r}$ excitation scheme (with $\ket{ns}$ corresponding to the electronic ground state) in the alkali atoms~\cite{Xu2015}. However, the experimental challenge will be to isolate a three-level system without spuriously populating additional intermediate or Rydberg states. Interestingly, the sign of the three-level dressed state potential is independent of the sign of the bare Rydberg-Rydberg interactions, opening the possibility to study purely repulsive anisotropic interactions, thereby minimizing losses or dipolar relaxation~\cite{Hensler2003}. Another interesting feature of three-level dressing is the possibility to introduce and control density dependent dissipation, which could be advantageous for studying, for example, non-equilibrium superfluidity in coupled quantum fluids of matter and light with strong and tuneable interactions~\cite{Carusotto2013}.

\acknowledgements{We acknowledge valuable discussions with Christopher Gaul and contributions to early theoretical calculations by Hanna Schempp. This work is supported in part by the Heidelberg Center for Quantum Dynamics, the European Union H2020 FET Proactive project RySQ (grant N. 640378), the 7th Framework Programme Initial Training Network (COHERENCE) and the Deutsche Forschungsgemeinschaft under WH141/1-1. SH acknowledges support by the Carl-Zeiss foundation. }

\bibliography{libraryNilsVersion}

\begin{thebibliography}{60}%
\makeatletter
\providecommand \@ifxundefined [1]{%
 \@ifx{#1\undefined}
}%
\providecommand \@ifnum [1]{%
 \ifnum #1\expandafter \@firstoftwo
 \else \expandafter \@secondoftwo
 \fi
}%
\providecommand \@ifx [1]{%
 \ifx #1\expandafter \@firstoftwo
 \else \expandafter \@secondoftwo
 \fi
}%
\providecommand \natexlab [1]{#1}%
\providecommand \enquote  [1]{``#1''}%
\providecommand \bibnamefont  [1]{#1}%
\providecommand \bibfnamefont [1]{#1}%
\providecommand \citenamefont [1]{#1}%
\providecommand \href@noop [0]{\@secondoftwo}%
\providecommand \href [0]{\begingroup \@sanitize@url \@href}%
\providecommand \@href[1]{\@@startlink{#1}\@@href}%
\providecommand \@@href[1]{\endgroup#1\@@endlink}%
\providecommand \@sanitize@url [0]{\catcode `\\12\catcode `\$12\catcode
  `\&12\catcode `\#12\catcode `\^12\catcode `\_12\catcode `\%12\relax}%
\providecommand \@@startlink[1]{}%
\providecommand \@@endlink[0]{}%
\providecommand \url  [0]{\begingroup\@sanitize@url \@url }%
\providecommand \@url [1]{\endgroup\@href {#1}{\urlprefix }}%
\providecommand \urlprefix  [0]{URL }%
\providecommand \Eprint [0]{\href }%
\providecommand \doibase [0]{http://dx.doi.org/}%
\providecommand \selectlanguage [0]{\@gobble}%
\providecommand \bibinfo  [0]{\@secondoftwo}%
\providecommand \bibfield  [0]{\@secondoftwo}%
\providecommand \translation [1]{[#1]}%
\providecommand \BibitemOpen [0]{}%
\providecommand \bibitemStop [0]{}%
\providecommand \bibitemNoStop [0]{.\EOS\space}%
\providecommand \EOS [0]{\spacefactor3000\relax}%
\providecommand \BibitemShut  [1]{\csname bibitem#1\endcsname}%
\let\auto@bib@innerbib\@empty
%</preamble>
\bibitem [{\citenamefont {Anderson}\ \emph {et~al.}(1995)\citenamefont
  {Anderson}, \citenamefont {Ensher}, \citenamefont {Matthews}, \citenamefont
  {Wieman},\ and\ \citenamefont {Cornell}}]{Anderson1995}%
  \BibitemOpen
  \bibfield  {author} {\bibinfo {author} {\bibfnamefont {M.~H.}\ \bibnamefont
  {Anderson}}, \bibinfo {author} {\bibfnamefont {J.~R.}\ \bibnamefont
  {Ensher}}, \bibinfo {author} {\bibfnamefont {M.~R.}\ \bibnamefont
  {Matthews}}, \bibinfo {author} {\bibfnamefont {C.~E.}\ \bibnamefont
  {Wieman}}, \ and\ \bibinfo {author} {\bibfnamefont {E.~A.}\ \bibnamefont
  {Cornell}},\ }\href {\doibase 10.1126/science.269.5221.198} {\bibfield
  {journal} {\bibinfo  {journal} {Science}\ }\textbf {\bibinfo {volume}
  {269}},\ \bibinfo {pages} {198} (\bibinfo {year} {1995})}\BibitemShut
  {NoStop}%
\bibitem [{\citenamefont {Davis}\ \emph {et~al.}(1995)\citenamefont {Davis},
  \citenamefont {Mewes}, \citenamefont {Andrews}, \citenamefont {{Van Druten}},
  \citenamefont {Durfee}, \citenamefont {Kurn},\ and\ \citenamefont
  {Ketterle}}]{Davis1995}%
  \BibitemOpen
  \bibfield  {author} {\bibinfo {author} {\bibfnamefont {K.~B.}\ \bibnamefont
  {Davis}}, \bibinfo {author} {\bibfnamefont {M.~O.}\ \bibnamefont {Mewes}},
  \bibinfo {author} {\bibfnamefont {M.~R.}\ \bibnamefont {Andrews}}, \bibinfo
  {author} {\bibfnamefont {N.~J.}\ \bibnamefont {{Van Druten}}}, \bibinfo
  {author} {\bibfnamefont {D.~S.}\ \bibnamefont {Durfee}}, \bibinfo {author}
  {\bibfnamefont {D.~M.}\ \bibnamefont {Kurn}}, \ and\ \bibinfo {author}
  {\bibfnamefont {W.}~\bibnamefont {Ketterle}},\ }\href {\doibase
  10.1103/PhysRevLett.75.3969} {\bibfield  {journal} {\bibinfo  {journal}
  {Phys. Rev. Lett.}\ }\textbf {\bibinfo {volume} {75}},\ \bibinfo {pages}
  {3969} (\bibinfo {year} {1995})}\BibitemShut {NoStop}%
\bibitem [{\citenamefont {Bradley}\ \emph {et~al.}(1995)\citenamefont
  {Bradley}, \citenamefont {Sackett}, \citenamefont {Tollett},\ and\
  \citenamefont {Hulet}}]{Bradley1995}%
  \BibitemOpen
  \bibfield  {author} {\bibinfo {author} {\bibfnamefont {C.~C.}\ \bibnamefont
  {Bradley}}, \bibinfo {author} {\bibfnamefont {C.~A.}\ \bibnamefont
  {Sackett}}, \bibinfo {author} {\bibfnamefont {J.~J.}\ \bibnamefont
  {Tollett}}, \ and\ \bibinfo {author} {\bibfnamefont {R.~G.}\ \bibnamefont
  {Hulet}},\ }\href {\doibase 10.1103/PhysRevLett.75.1687} {\bibfield
  {journal} {\bibinfo  {journal} {Phys. Rev. Lett.}\ }\textbf {\bibinfo
  {volume} {75}},\ \bibinfo {pages} {1687} (\bibinfo {year}
  {1995})}\BibitemShut {NoStop}%
\bibitem [{\citenamefont {Greiner}\ \emph {et~al.}(2003)\citenamefont
  {Greiner}, \citenamefont {Regal},\ and\ \citenamefont {Jin}}]{Greiner2003}%
  \BibitemOpen
  \bibfield  {author} {\bibinfo {author} {\bibfnamefont {M.}~\bibnamefont
  {Greiner}}, \bibinfo {author} {\bibfnamefont {C.~A.}\ \bibnamefont {Regal}},
  \ and\ \bibinfo {author} {\bibfnamefont {D.~S.}\ \bibnamefont {Jin}},\ }\href
  {\doibase 10.1038/nature02199} {\bibfield  {journal} {\bibinfo  {journal}
  {Nature}\ }\textbf {\bibinfo {volume} {426}},\ \bibinfo {pages} {537}
  (\bibinfo {year} {2003})}\BibitemShut {NoStop}%
\bibitem [{\citenamefont {Jochim}\ \emph {et~al.}(2003)\citenamefont {Jochim},
  \citenamefont {Bartenstein}, \citenamefont {Altmeyer}, \citenamefont {Hendl},
  \citenamefont {Riedl}, \citenamefont {Chin}, \citenamefont {Denschlag},\ and\
  \citenamefont {Grimm}}]{Jochim2003}%
  \BibitemOpen
  \bibfield  {author} {\bibinfo {author} {\bibfnamefont {S.}~\bibnamefont
  {Jochim}}, \bibinfo {author} {\bibfnamefont {M.}~\bibnamefont {Bartenstein}},
  \bibinfo {author} {\bibfnamefont {A.}~\bibnamefont {Altmeyer}}, \bibinfo
  {author} {\bibfnamefont {G.}~\bibnamefont {Hendl}}, \bibinfo {author}
  {\bibfnamefont {S.}~\bibnamefont {Riedl}}, \bibinfo {author} {\bibfnamefont
  {C.}~\bibnamefont {Chin}}, \bibinfo {author} {\bibfnamefont {J.~H.}\
  \bibnamefont {Denschlag}}, \ and\ \bibinfo {author} {\bibfnamefont
  {R.}~\bibnamefont {Grimm}},\ }\href {\doibase 10.1126/science.1093280}
  {\bibfield  {journal} {\bibinfo  {journal} {Science}\ }\textbf {\bibinfo
  {volume} {302}},\ \bibinfo {pages} {2101} (\bibinfo {year}
  {2003})}\BibitemShut {NoStop}%
\bibitem [{\citenamefont {Zwierlein}\ \emph {et~al.}(2003)\citenamefont
  {Zwierlein}, \citenamefont {Stan}, \citenamefont {Schunck}, \citenamefont
  {Raupach}, \citenamefont {Gupta}, \citenamefont {Hadzibabic},\ and\
  \citenamefont {Ketterle}}]{Zwierlein2003}%
  \BibitemOpen
  \bibfield  {author} {\bibinfo {author} {\bibfnamefont {M.~W.}\ \bibnamefont
  {Zwierlein}}, \bibinfo {author} {\bibfnamefont {C.~A.}\ \bibnamefont {Stan}},
  \bibinfo {author} {\bibfnamefont {C.~H.}\ \bibnamefont {Schunck}}, \bibinfo
  {author} {\bibfnamefont {S.~M.~F.}\ \bibnamefont {Raupach}}, \bibinfo
  {author} {\bibfnamefont {S.}~\bibnamefont {Gupta}}, \bibinfo {author}
  {\bibfnamefont {Z.}~\bibnamefont {Hadzibabic}}, \ and\ \bibinfo {author}
  {\bibfnamefont {W.}~\bibnamefont {Ketterle}},\ }\href {\doibase
  10.1103/PhysRevLett.91.250401} {\bibfield  {journal} {\bibinfo  {journal}
  {Phys. Rev. Lett.}\ }\textbf {\bibinfo {volume} {91}},\ \bibinfo {pages}
  {250401} (\bibinfo {year} {2003})}\BibitemShut {NoStop}%
\bibitem [{\citenamefont {Regal}\ \emph {et~al.}(2004)\citenamefont {Regal},
  \citenamefont {Greiner},\ and\ \citenamefont {Jin}}]{Regal2004}%
  \BibitemOpen
  \bibfield  {author} {\bibinfo {author} {\bibfnamefont {C.~A.}\ \bibnamefont
  {Regal}}, \bibinfo {author} {\bibfnamefont {M.}~\bibnamefont {Greiner}}, \
  and\ \bibinfo {author} {\bibfnamefont {D.~S.}\ \bibnamefont {Jin}},\ }\href
  {\doibase 10.1103/PhysRevLett.92.040403} {\bibfield  {journal} {\bibinfo
  {journal} {Phys. Rev. Lett.}\ }\textbf {\bibinfo {volume} {92}},\ \bibinfo
  {pages} {040403} (\bibinfo {year} {2004})}\BibitemShut {NoStop}%
\bibitem [{\citenamefont {Zwierlein}\ \emph {et~al.}(2004)\citenamefont
  {Zwierlein}, \citenamefont {Stan}, \citenamefont {Schunck}, \citenamefont
  {Raupach}, \citenamefont {Kerman},\ and\ \citenamefont
  {Ketterle}}]{Zwierlein2004}%
  \BibitemOpen
  \bibfield  {author} {\bibinfo {author} {\bibfnamefont {M.~W.}\ \bibnamefont
  {Zwierlein}}, \bibinfo {author} {\bibfnamefont {C.~A.}\ \bibnamefont {Stan}},
  \bibinfo {author} {\bibfnamefont {C.~H.}\ \bibnamefont {Schunck}}, \bibinfo
  {author} {\bibfnamefont {S.~M.~F.}\ \bibnamefont {Raupach}}, \bibinfo
  {author} {\bibfnamefont {A.~J.}\ \bibnamefont {Kerman}}, \ and\ \bibinfo
  {author} {\bibfnamefont {W.}~\bibnamefont {Ketterle}},\ }\href {\doibase
  10.1103/PhysRevLett.92.120403} {\bibfield  {journal} {\bibinfo  {journal}
  {Phys. Rev. Lett.}\ }\textbf {\bibinfo {volume} {92}},\ \bibinfo {pages}
  {120403} (\bibinfo {year} {2004})}\BibitemShut {NoStop}%
\bibitem [{\citenamefont {Kinast}\ \emph {et~al.}(2004)\citenamefont {Kinast},
  \citenamefont {Hemmer}, \citenamefont {Gehm}, \citenamefont {Turlapov},\ and\
  \citenamefont {Thomas}}]{Kinast2004}%
  \BibitemOpen
  \bibfield  {author} {\bibinfo {author} {\bibfnamefont {J.}~\bibnamefont
  {Kinast}}, \bibinfo {author} {\bibfnamefont {S.~L.}\ \bibnamefont {Hemmer}},
  \bibinfo {author} {\bibfnamefont {M.~E.}\ \bibnamefont {Gehm}}, \bibinfo
  {author} {\bibfnamefont {A.}~\bibnamefont {Turlapov}}, \ and\ \bibinfo
  {author} {\bibfnamefont {J.~E.}\ \bibnamefont {Thomas}},\ }\href {\doibase
  10.1103/PhysRevLett.92.150402} {\bibfield  {journal} {\bibinfo  {journal}
  {Phys. Rev. Lett.}\ }\textbf {\bibinfo {volume} {92}},\ \bibinfo {pages}
  {150402} (\bibinfo {year} {2004})}\BibitemShut {NoStop}%
\bibitem [{\citenamefont {Bourdel}\ \emph {et~al.}(2004)\citenamefont
  {Bourdel}, \citenamefont {Khaykovich}, \citenamefont {Cubizolles},
  \citenamefont {Zhang}, \citenamefont {Chevy}, \citenamefont {Teichmann},
  \citenamefont {Tarruell}, \citenamefont {Kokkelmans},\ and\ \citenamefont
  {Salomon}}]{Bourdel2004}%
  \BibitemOpen
  \bibfield  {author} {\bibinfo {author} {\bibfnamefont {T.}~\bibnamefont
  {Bourdel}}, \bibinfo {author} {\bibfnamefont {L.}~\bibnamefont {Khaykovich}},
  \bibinfo {author} {\bibfnamefont {J.}~\bibnamefont {Cubizolles}}, \bibinfo
  {author} {\bibfnamefont {J.}~\bibnamefont {Zhang}}, \bibinfo {author}
  {\bibfnamefont {F.}~\bibnamefont {Chevy}}, \bibinfo {author} {\bibfnamefont
  {M.}~\bibnamefont {Teichmann}}, \bibinfo {author} {\bibfnamefont
  {L.}~\bibnamefont {Tarruell}}, \bibinfo {author} {\bibfnamefont {S.~J. J.
  M.~F.}\ \bibnamefont {Kokkelmans}}, \ and\ \bibinfo {author} {\bibfnamefont
  {C.}~\bibnamefont {Salomon}},\ }\href {\doibase
  10.1103/PhysRevLett.93.050401} {\bibfield  {journal} {\bibinfo  {journal}
  {Phys. Rev. Lett.}\ }\textbf {\bibinfo {volume} {93}},\ \bibinfo {pages}
  {050401} (\bibinfo {year} {2004})}\BibitemShut {NoStop}%
\bibitem [{\citenamefont {Chin}\ \emph {et~al.}(2004)\citenamefont {Chin},
  \citenamefont {Bartenstein}, \citenamefont {Altmeyer}, \citenamefont {Riedl},
  \citenamefont {Jochim}, \citenamefont {Denschlag},\ and\ \citenamefont
  {Grimm}}]{Chin2004}%
  \BibitemOpen
  \bibfield  {author} {\bibinfo {author} {\bibfnamefont {C.}~\bibnamefont
  {Chin}}, \bibinfo {author} {\bibfnamefont {M.}~\bibnamefont {Bartenstein}},
  \bibinfo {author} {\bibfnamefont {A.}~\bibnamefont {Altmeyer}}, \bibinfo
  {author} {\bibfnamefont {S.}~\bibnamefont {Riedl}}, \bibinfo {author}
  {\bibfnamefont {S.}~\bibnamefont {Jochim}}, \bibinfo {author} {\bibfnamefont
  {J.~H.}\ \bibnamefont {Denschlag}}, \ and\ \bibinfo {author} {\bibfnamefont
  {R.}~\bibnamefont {Grimm}},\ }\href {\doibase 10.1126/science.1100818}
  {\bibfield  {journal} {\bibinfo  {journal} {Science}\ }\textbf {\bibinfo
  {volume} {305}},\ \bibinfo {pages} {1128} (\bibinfo {year}
  {2004})}\BibitemShut {NoStop}%
\bibitem [{\citenamefont {Dagotto}(2005)}]{Dagotto2005}%
  \BibitemOpen
  \bibfield  {author} {\bibinfo {author} {\bibfnamefont {E.}~\bibnamefont
  {Dagotto}},\ }\href {\doibase 10.1126/science.1107559} {\bibfield  {journal}
  {\bibinfo  {journal} {Science}\ }\textbf {\bibinfo {volume} {309}},\ \bibinfo
  {pages} {257} (\bibinfo {year} {2005})}\BibitemShut {NoStop}%
\bibitem [{\citenamefont {de~Gennes}\ and\ \citenamefont
  {Prost}(1993)}]{Gennes1993}%
  \BibitemOpen
  \bibfield  {author} {\bibinfo {author} {\bibfnamefont {P.~G.}\ \bibnamefont
  {de~Gennes}}\ and\ \bibinfo {author} {\bibfnamefont {J.}~\bibnamefont
  {Prost}},\ }\href@noop {} {\emph {\bibinfo {title} {{The Physics of Liquid
  Crystals}}}}\ (\bibinfo  {publisher} {Oxford University Press, International
  Series of Monographs on Physics 83},\ \bibinfo {year} {1993})\BibitemShut
  {NoStop}%
\bibitem [{\citenamefont {Lahaye}\ \emph {et~al.}(2009)\citenamefont {Lahaye},
  \citenamefont {Menotti}, \citenamefont {Santos}, \citenamefont {Lewenstein},\
  and\ \citenamefont {Pfau}}]{Lahaye2009}%
  \BibitemOpen
  \bibfield  {author} {\bibinfo {author} {\bibfnamefont {T.}~\bibnamefont
  {Lahaye}}, \bibinfo {author} {\bibfnamefont {C.}~\bibnamefont {Menotti}},
  \bibinfo {author} {\bibfnamefont {L.}~\bibnamefont {Santos}}, \bibinfo
  {author} {\bibfnamefont {M.}~\bibnamefont {Lewenstein}}, \ and\ \bibinfo
  {author} {\bibfnamefont {T.}~\bibnamefont {Pfau}},\ }\href {\doibase
  10.1088/0034-4885/72/12/126401} {\bibfield  {journal} {\bibinfo  {journal}
  {Rep. Prog. Phys.}\ }\textbf {\bibinfo {volume} {72}},\ \bibinfo {pages} {71}
  (\bibinfo {year} {2009})}\BibitemShut {NoStop}%
\bibitem [{\citenamefont {Stuhler}\ \emph {et~al.}(2005)\citenamefont
  {Stuhler}, \citenamefont {Griesmaier}, \citenamefont {Koch}, \citenamefont
  {Fattori}, \citenamefont {Pfau}, \citenamefont {Giovanazzi}, \citenamefont
  {Pedri},\ and\ \citenamefont {Santos}}]{Stuhler2005}%
  \BibitemOpen
  \bibfield  {author} {\bibinfo {author} {\bibfnamefont {J.}~\bibnamefont
  {Stuhler}}, \bibinfo {author} {\bibfnamefont {A.}~\bibnamefont {Griesmaier}},
  \bibinfo {author} {\bibfnamefont {T.}~\bibnamefont {Koch}}, \bibinfo {author}
  {\bibfnamefont {M.}~\bibnamefont {Fattori}}, \bibinfo {author} {\bibfnamefont
  {T.}~\bibnamefont {Pfau}}, \bibinfo {author} {\bibfnamefont {S.}~\bibnamefont
  {Giovanazzi}}, \bibinfo {author} {\bibfnamefont {P.}~\bibnamefont {Pedri}}, \
  and\ \bibinfo {author} {\bibfnamefont {L.}~\bibnamefont {Santos}},\ }\href
  {\doibase 10.1103/PhysRevLett.95.150406} {\bibfield  {journal} {\bibinfo
  {journal} {Phys. Rev. Lett.}\ }\textbf {\bibinfo {volume} {95}},\ \bibinfo
  {pages} {150406} (\bibinfo {year} {2005})}\BibitemShut {NoStop}%
\bibitem [{\citenamefont {Aikawa}\ \emph {et~al.}(2012)\citenamefont {Aikawa},
  \citenamefont {Frisch}, \citenamefont {Mark}, \citenamefont {Baier},
  \citenamefont {Rietzler}, \citenamefont {Grimm},\ and\ \citenamefont
  {Ferlaino}}]{Aikawa2012}%
  \BibitemOpen
  \bibfield  {author} {\bibinfo {author} {\bibfnamefont {K.}~\bibnamefont
  {Aikawa}}, \bibinfo {author} {\bibfnamefont {A.}~\bibnamefont {Frisch}},
  \bibinfo {author} {\bibfnamefont {M.}~\bibnamefont {Mark}}, \bibinfo {author}
  {\bibfnamefont {S.}~\bibnamefont {Baier}}, \bibinfo {author} {\bibfnamefont
  {A.}~\bibnamefont {Rietzler}}, \bibinfo {author} {\bibfnamefont
  {R.}~\bibnamefont {Grimm}}, \ and\ \bibinfo {author} {\bibfnamefont
  {F.}~\bibnamefont {Ferlaino}},\ }\href {\doibase
  10.1103/PhysRevLett.108.210401} {\bibfield  {journal} {\bibinfo  {journal}
  {Phys. Rev. Lett.}\ }\textbf {\bibinfo {volume} {108}},\ \bibinfo {pages}
  {210401} (\bibinfo {year} {2012})}\BibitemShut {NoStop}%
\bibitem [{\citenamefont {Lu}\ \emph {et~al.}(2012)\citenamefont {Lu},
  \citenamefont {Burdick},\ and\ \citenamefont {Lev}}]{Lu2012}%
  \BibitemOpen
  \bibfield  {author} {\bibinfo {author} {\bibfnamefont {M.}~\bibnamefont
  {Lu}}, \bibinfo {author} {\bibfnamefont {N.~Q.}\ \bibnamefont {Burdick}}, \
  and\ \bibinfo {author} {\bibfnamefont {B.~L.}\ \bibnamefont {Lev}},\ }\href
  {\doibase 10.1103/PhysRevLett.108.215301} {\bibfield  {journal} {\bibinfo
  {journal} {Phys. Rev. Lett.}\ }\textbf {\bibinfo {volume} {108}},\ \bibinfo
  {pages} {215301} (\bibinfo {year} {2012})}\BibitemShut {NoStop}%
\bibitem [{\citenamefont {Deiglmayr}\ \emph {et~al.}(2010)\citenamefont
  {Deiglmayr}, \citenamefont {Grochola}, \citenamefont {Repp}, \citenamefont
  {Dulieu}, \citenamefont {Wester},\ and\ \citenamefont
  {Weidem\"{u}ller}}]{Deiglmayr2010}%
  \BibitemOpen
  \bibfield  {author} {\bibinfo {author} {\bibfnamefont {J.}~\bibnamefont
  {Deiglmayr}}, \bibinfo {author} {\bibfnamefont {A.}~\bibnamefont {Grochola}},
  \bibinfo {author} {\bibfnamefont {M.}~\bibnamefont {Repp}}, \bibinfo {author}
  {\bibfnamefont {O.}~\bibnamefont {Dulieu}}, \bibinfo {author} {\bibfnamefont
  {R.}~\bibnamefont {Wester}}, \ and\ \bibinfo {author} {\bibfnamefont
  {M.}~\bibnamefont {Weidem\"{u}ller}},\ }\href
  {http://pra.aps.org/pdf/PRA/v82/i3/e032503} {\bibfield  {journal} {\bibinfo
  {journal} {Phys. Rev. A}\ }\textbf {\bibinfo {volume} {82}},\ \bibinfo
  {pages} {32503} (\bibinfo {year} {2010})}\BibitemShut {NoStop}%
\bibitem [{\citenamefont {Ni}\ \emph {et~al.}(2008)\citenamefont {Ni},
  \citenamefont {Ospelkaus}, \citenamefont {de~Miranda}, \citenamefont {Pe'er},
  \citenamefont {Neyenhuis}, \citenamefont {Zirbel}, \citenamefont
  {Kotochigova}, \citenamefont {Julienne}, \citenamefont {Jin},\ and\
  \citenamefont {Ye}}]{Ni2008}%
  \BibitemOpen
  \bibfield  {author} {\bibinfo {author} {\bibfnamefont {K.-K.}\ \bibnamefont
  {Ni}}, \bibinfo {author} {\bibfnamefont {S.}~\bibnamefont {Ospelkaus}},
  \bibinfo {author} {\bibfnamefont {M.}~\bibnamefont {de~Miranda}}, \bibinfo
  {author} {\bibfnamefont {A.}~\bibnamefont {Pe'er}}, \bibinfo {author}
  {\bibfnamefont {B.}~\bibnamefont {Neyenhuis}}, \bibinfo {author}
  {\bibfnamefont {J.}~\bibnamefont {Zirbel}}, \bibinfo {author} {\bibfnamefont
  {S.}~\bibnamefont {Kotochigova}}, \bibinfo {author} {\bibfnamefont
  {P.}~\bibnamefont {Julienne}}, \bibinfo {author} {\bibfnamefont
  {D.}~\bibnamefont {Jin}}, \ and\ \bibinfo {author} {\bibfnamefont
  {J.}~\bibnamefont {Ye}},\ }\href@noop {} {\bibfield  {journal} {\bibinfo
  {journal} {Science}\ }\textbf {\bibinfo {volume} {322}},\ \bibinfo {pages}
  {231} (\bibinfo {year} {2008})}\BibitemShut {NoStop}%
\bibitem [{\citenamefont {Takekoshi}\ \emph {et~al.}(2014)\citenamefont
  {Takekoshi}, \citenamefont {Reichs{\"o}llner}, \citenamefont {Schindewolf},
  \citenamefont {Hutson}, \citenamefont {Le~Sueur}, \citenamefont {Dulieu},
  \citenamefont {Ferlaino}, \citenamefont {Grimm},\ and\ \citenamefont
  {N{\"a}gerl}}]{Takekoshi2014}%
  \BibitemOpen
  \bibfield  {author} {\bibinfo {author} {\bibfnamefont {T.}~\bibnamefont
  {Takekoshi}}, \bibinfo {author} {\bibfnamefont {L.}~\bibnamefont
  {Reichs{\"o}llner}}, \bibinfo {author} {\bibfnamefont {A.}~\bibnamefont
  {Schindewolf}}, \bibinfo {author} {\bibfnamefont {J.~M.}\ \bibnamefont
  {Hutson}}, \bibinfo {author} {\bibfnamefont {C.~R.}\ \bibnamefont
  {Le~Sueur}}, \bibinfo {author} {\bibfnamefont {O.}~\bibnamefont {Dulieu}},
  \bibinfo {author} {\bibfnamefont {F.}~\bibnamefont {Ferlaino}}, \bibinfo
  {author} {\bibfnamefont {R.}~\bibnamefont {Grimm}}, \ and\ \bibinfo {author}
  {\bibfnamefont {H.-C.}\ \bibnamefont {N{\"a}gerl}},\ }\href@noop {}
  {\bibfield  {journal} {\bibinfo  {journal} {Phys. Rev. Lett.}\ }\textbf
  {\bibinfo {volume} {113}},\ \bibinfo {pages} {205301} (\bibinfo {year}
  {2014})}\BibitemShut {NoStop}%
\bibitem [{\citenamefont {Lahaye}\ \emph {et~al.}(2008)\citenamefont {Lahaye},
  \citenamefont {Metz}, \citenamefont {Fr\"{o}hlich}, \citenamefont {Koch},
  \citenamefont {Meister}, \citenamefont {Griesmaier}, \citenamefont {Pfau},
  \citenamefont {Saito}, \citenamefont {Kawaguchi},\ and\ \citenamefont
  {Ueda}}]{Lahaye2008}%
  \BibitemOpen
  \bibfield  {author} {\bibinfo {author} {\bibfnamefont {T.}~\bibnamefont
  {Lahaye}}, \bibinfo {author} {\bibfnamefont {J.}~\bibnamefont {Metz}},
  \bibinfo {author} {\bibfnamefont {B.}~\bibnamefont {Fr\"{o}hlich}}, \bibinfo
  {author} {\bibfnamefont {T.}~\bibnamefont {Koch}}, \bibinfo {author}
  {\bibfnamefont {M.}~\bibnamefont {Meister}}, \bibinfo {author} {\bibfnamefont
  {A.}~\bibnamefont {Griesmaier}}, \bibinfo {author} {\bibfnamefont
  {T.}~\bibnamefont {Pfau}}, \bibinfo {author} {\bibfnamefont {H.}~\bibnamefont
  {Saito}}, \bibinfo {author} {\bibfnamefont {Y.}~\bibnamefont {Kawaguchi}}, \
  and\ \bibinfo {author} {\bibfnamefont {M.}~\bibnamefont {Ueda}},\ }\href
  {\doibase 10.1103/PhysRevLett.101.080401} {\bibfield  {journal} {\bibinfo
  {journal} {Phys. Rev. Lett.}\ }\textbf {\bibinfo {volume} {101}},\ \bibinfo
  {pages} {080401} (\bibinfo {year} {2008})}\BibitemShut {NoStop}%
\bibitem [{\citenamefont {Sadler}\ \emph {et~al.}(2006)\citenamefont {Sadler},
  \citenamefont {Higbie}, \citenamefont {Leslie}, \citenamefont
  {Vengalattore},\ and\ \citenamefont {Stamper-Kurn}}]{Sadler2006}%
  \BibitemOpen
  \bibfield  {author} {\bibinfo {author} {\bibfnamefont {L.~E.}\ \bibnamefont
  {Sadler}}, \bibinfo {author} {\bibfnamefont {J.~M.}\ \bibnamefont {Higbie}},
  \bibinfo {author} {\bibfnamefont {S.~R.}\ \bibnamefont {Leslie}}, \bibinfo
  {author} {\bibfnamefont {M.}~\bibnamefont {Vengalattore}}, \ and\ \bibinfo
  {author} {\bibfnamefont {D.~M.}\ \bibnamefont {Stamper-Kurn}},\ }\href
  {\doibase 10.1038/nature05094} {\bibfield  {journal} {\bibinfo  {journal}
  {Nature}\ }\textbf {\bibinfo {volume} {443}},\ \bibinfo {pages} {312}
  (\bibinfo {year} {2006})}\BibitemShut {NoStop}%
\bibitem [{\citenamefont {Pasquiou}\ \emph {et~al.}(2011)\citenamefont
  {Pasquiou}, \citenamefont {Mar\'{e}chal}, \citenamefont {Bismut},
  \citenamefont {Pedri}, \citenamefont {Vernac}, \citenamefont {Gorceix},\ and\
  \citenamefont {Laburthe-Tolra}}]{Pasquiou2011}%
  \BibitemOpen
  \bibfield  {author} {\bibinfo {author} {\bibfnamefont {B.}~\bibnamefont
  {Pasquiou}}, \bibinfo {author} {\bibfnamefont {E.}~\bibnamefont
  {Mar\'{e}chal}}, \bibinfo {author} {\bibfnamefont {G.}~\bibnamefont
  {Bismut}}, \bibinfo {author} {\bibfnamefont {P.}~\bibnamefont {Pedri}},
  \bibinfo {author} {\bibfnamefont {L.}~\bibnamefont {Vernac}}, \bibinfo
  {author} {\bibfnamefont {O.}~\bibnamefont {Gorceix}}, \ and\ \bibinfo
  {author} {\bibfnamefont {B.}~\bibnamefont {Laburthe-Tolra}},\ }\href
  {\doibase 10.1103/PhysRevLett.106.255303} {\bibfield  {journal} {\bibinfo
  {journal} {Phys. Rev. Lett.}\ }\textbf {\bibinfo {volume} {106}},\ \bibinfo
  {pages} {1} (\bibinfo {year} {2011})}\BibitemShut {NoStop}%
\bibitem [{\citenamefont {Aikawa}\ \emph {et~al.}(2014)\citenamefont {Aikawa},
  \citenamefont {Frisch}, \citenamefont {Mark}, \citenamefont {Baier},
  \citenamefont {Grimm},\ and\ \citenamefont {Ferlaino}}]{Aikawa2014b}%
  \BibitemOpen
  \bibfield  {author} {\bibinfo {author} {\bibfnamefont {K.}~\bibnamefont
  {Aikawa}}, \bibinfo {author} {\bibfnamefont {A.}~\bibnamefont {Frisch}},
  \bibinfo {author} {\bibfnamefont {M.}~\bibnamefont {Mark}}, \bibinfo {author}
  {\bibfnamefont {S.}~\bibnamefont {Baier}}, \bibinfo {author} {\bibfnamefont
  {R.}~\bibnamefont {Grimm}}, \ and\ \bibinfo {author} {\bibfnamefont
  {F.}~\bibnamefont {Ferlaino}},\ }\href {\doibase
  10.1103/PhysRevLett.112.010404} {\bibfield  {journal} {\bibinfo  {journal}
  {Phys. Rev. Lett.}\ }\textbf {\bibinfo {volume} {112}},\ \bibinfo {pages}
  {010404} (\bibinfo {year} {2014})}\BibitemShut {NoStop}%
\bibitem [{\citenamefont {Kadau}\ \emph {et~al.}(2015)\citenamefont {Kadau},
  \citenamefont {Schmitt}, \citenamefont {Wenzel}, \citenamefont {Wink},
  \citenamefont {Maier}, \citenamefont {Ferrier-Barbut},\ and\ \citenamefont
  {Pfau}}]{Kadau2015}%
  \BibitemOpen
  \bibfield  {author} {\bibinfo {author} {\bibfnamefont {H.}~\bibnamefont
  {Kadau}}, \bibinfo {author} {\bibfnamefont {M.}~\bibnamefont {Schmitt}},
  \bibinfo {author} {\bibfnamefont {M.}~\bibnamefont {Wenzel}}, \bibinfo
  {author} {\bibfnamefont {C.}~\bibnamefont {Wink}}, \bibinfo {author}
  {\bibfnamefont {T.}~\bibnamefont {Maier}}, \bibinfo {author} {\bibfnamefont
  {I.}~\bibnamefont {Ferrier-Barbut}}, \ and\ \bibinfo {author} {\bibfnamefont
  {T.}~\bibnamefont {Pfau}},\ }\href@noop {} {\bibfield  {journal} {\bibinfo
  {journal} {arXiv:1508.05007}\ } (\bibinfo {year} {2015})}\BibitemShut
  {NoStop}%
\bibitem [{\citenamefont {de~Paz}\ \emph {et~al.}(2013)\citenamefont {de~Paz},
  \citenamefont {Sharma}, \citenamefont {Chotia}, \citenamefont {Mar\'{e}chal},
  \citenamefont {Huckans}, \citenamefont {Pedri}, \citenamefont {Santos},
  \citenamefont {Gorceix}, \citenamefont {Vernac},\ and\ \citenamefont
  {Laburthe-Tolra}}]{DePaz2013}%
  \BibitemOpen
  \bibfield  {author} {\bibinfo {author} {\bibfnamefont {A.}~\bibnamefont
  {de~Paz}}, \bibinfo {author} {\bibfnamefont {A.}~\bibnamefont {Sharma}},
  \bibinfo {author} {\bibfnamefont {A.}~\bibnamefont {Chotia}}, \bibinfo
  {author} {\bibfnamefont {E.}~\bibnamefont {Mar\'{e}chal}}, \bibinfo {author}
  {\bibfnamefont {J.~H.}\ \bibnamefont {Huckans}}, \bibinfo {author}
  {\bibfnamefont {P.}~\bibnamefont {Pedri}}, \bibinfo {author} {\bibfnamefont
  {L.}~\bibnamefont {Santos}}, \bibinfo {author} {\bibfnamefont
  {O.}~\bibnamefont {Gorceix}}, \bibinfo {author} {\bibfnamefont
  {L.}~\bibnamefont {Vernac}}, \ and\ \bibinfo {author} {\bibfnamefont
  {B.}~\bibnamefont {Laburthe-Tolra}},\ }\href {\doibase
  10.1103/PhysRevLett.111.185305} {\bibfield  {journal} {\bibinfo  {journal}
  {Phys. Rev. Lett.}\ }\textbf {\bibinfo {volume} {111}},\ \bibinfo {pages}
  {185305} (\bibinfo {year} {2013})}\BibitemShut {NoStop}%
\bibitem [{\citenamefont {Baier}\ \emph {et~al.}(2015)\citenamefont {Baier},
  \citenamefont {Mark}, \citenamefont {Petter}, \citenamefont {Aikawa},
  \citenamefont {Chomaz}, \citenamefont {Cai}, \citenamefont {Baranov},
  \citenamefont {Zoller},\ and\ \citenamefont {Ferlaino}}]{Baier2015}%
  \BibitemOpen
  \bibfield  {author} {\bibinfo {author} {\bibfnamefont {S.}~\bibnamefont
  {Baier}}, \bibinfo {author} {\bibfnamefont {M.~J.}\ \bibnamefont {Mark}},
  \bibinfo {author} {\bibfnamefont {D.}~\bibnamefont {Petter}}, \bibinfo
  {author} {\bibfnamefont {K.}~\bibnamefont {Aikawa}}, \bibinfo {author}
  {\bibfnamefont {L.}~\bibnamefont {Chomaz}}, \bibinfo {author} {\bibfnamefont
  {Z.}~\bibnamefont {Cai}}, \bibinfo {author} {\bibfnamefont {M.}~\bibnamefont
  {Baranov}}, \bibinfo {author} {\bibfnamefont {P.}~\bibnamefont {Zoller}}, \
  and\ \bibinfo {author} {\bibfnamefont {F.}~\bibnamefont {Ferlaino}},\ }\href
  {http://arxiv.org/abs/1507.03500} {\bibfield  {journal} {\bibinfo  {journal}
  {arXiv:1507.03500}\ } (\bibinfo {year} {2015})}\BibitemShut {NoStop}%
\bibitem [{\citenamefont {Yan}\ \emph {et~al.}(2013)\citenamefont {Yan},
  \citenamefont {Moses}, \citenamefont {Gadway}, \citenamefont {Covey},
  \citenamefont {Hazzard}, \citenamefont {Rey}, \citenamefont {Jin},\ and\
  \citenamefont {Ye}}]{Yan2013}%
  \BibitemOpen
  \bibfield  {author} {\bibinfo {author} {\bibfnamefont {B.}~\bibnamefont
  {Yan}}, \bibinfo {author} {\bibfnamefont {S.~A.}\ \bibnamefont {Moses}},
  \bibinfo {author} {\bibfnamefont {B.}~\bibnamefont {Gadway}}, \bibinfo
  {author} {\bibfnamefont {J.~P.}\ \bibnamefont {Covey}}, \bibinfo {author}
  {\bibfnamefont {K.~R.~A.}\ \bibnamefont {Hazzard}}, \bibinfo {author}
  {\bibfnamefont {A.~M.}\ \bibnamefont {Rey}}, \bibinfo {author} {\bibfnamefont
  {D.~S.}\ \bibnamefont {Jin}}, \ and\ \bibinfo {author} {\bibfnamefont
  {J.}~\bibnamefont {Ye}},\ }\href@noop {} {\bibfield  {journal} {\bibinfo
  {journal} {Nature}\ }\textbf {\bibinfo {volume} {501}},\ \bibinfo {pages}
  {521} (\bibinfo {year} {2013})}\BibitemShut {NoStop}%
\bibitem [{\citenamefont {Yao}\ \emph {et~al.}(2012)\citenamefont {Yao},
  \citenamefont {Laumann}, \citenamefont {Gorshkov}, \citenamefont {Bennett},
  \citenamefont {Demler}, \citenamefont {Zoller},\ and\ \citenamefont
  {Lukin}}]{Yao2012}%
  \BibitemOpen
  \bibfield  {author} {\bibinfo {author} {\bibfnamefont {N.~Y.}\ \bibnamefont
  {Yao}}, \bibinfo {author} {\bibfnamefont {C.~R.}\ \bibnamefont {Laumann}},
  \bibinfo {author} {\bibfnamefont {A.~V.}\ \bibnamefont {Gorshkov}}, \bibinfo
  {author} {\bibfnamefont {S.~D.}\ \bibnamefont {Bennett}}, \bibinfo {author}
  {\bibfnamefont {E.}~\bibnamefont {Demler}}, \bibinfo {author} {\bibfnamefont
  {P.}~\bibnamefont {Zoller}}, \ and\ \bibinfo {author} {\bibfnamefont {M.~D.}\
  \bibnamefont {Lukin}},\ }\href {\doibase 10.1103/PhysRevLett.109.266804}
  {\bibfield  {journal} {\bibinfo  {journal} {Phys. Rev. Lett.}\ }\textbf
  {\bibinfo {volume} {109}},\ \bibinfo {pages} {266804} (\bibinfo {year}
  {2012})}\BibitemShut {NoStop}%
\bibitem [{\citenamefont {Li}\ and\ \citenamefont {Sarma}(2015)}]{Li2015}%
  \BibitemOpen
  \bibfield  {author} {\bibinfo {author} {\bibfnamefont {X.}~\bibnamefont
  {Li}}\ and\ \bibinfo {author} {\bibfnamefont {S.~D.}\ \bibnamefont {Sarma}},\
  }\href {\doibase 10.1038/ncomms8137} {\bibfield  {journal} {\bibinfo
  {journal} {Nature Communications}\ }\textbf {\bibinfo {volume} {6}},\
  \bibinfo {pages} {7137} (\bibinfo {year} {2015})}\BibitemShut {NoStop}%
\bibitem [{\citenamefont {Peter}\ \emph {et~al.}(2014)\citenamefont {Peter},
  \citenamefont {Yao}, \citenamefont {Lang}, \citenamefont {Huber},
  \citenamefont {Lukin},\ and\ \citenamefont {B{\"{u}}chler}}]{Peter2014}%
  \BibitemOpen
  \bibfield  {author} {\bibinfo {author} {\bibfnamefont {D.}~\bibnamefont
  {Peter}}, \bibinfo {author} {\bibfnamefont {N.~Y.}\ \bibnamefont {Yao}},
  \bibinfo {author} {\bibfnamefont {N.}~\bibnamefont {Lang}}, \bibinfo {author}
  {\bibfnamefont {S.~D.}\ \bibnamefont {Huber}}, \bibinfo {author}
  {\bibfnamefont {M.~D.}\ \bibnamefont {Lukin}}, \ and\ \bibinfo {author}
  {\bibfnamefont {H.~P.}\ \bibnamefont {B{\"{u}}chler}},\ }\href {\doibase
  10.1103/PhysRevA.91.053617} {\bibfield  {journal} {\bibinfo  {journal} {Phys.
  Rev. A}\ }\textbf {\bibinfo {volume} {91}},\ \bibinfo {pages} {053617}
  (\bibinfo {year} {2014})}\BibitemShut {NoStop}%
\bibitem [{\citenamefont {van Bijnen}\ and\ \citenamefont
  {Pohl}(2015)}]{VanBijnen2014}%
  \BibitemOpen
  \bibfield  {author} {\bibinfo {author} {\bibfnamefont {R.~M.~W.}\
  \bibnamefont {van Bijnen}}\ and\ \bibinfo {author} {\bibfnamefont
  {T.}~\bibnamefont {Pohl}},\ }\href {\doibase 10.1103/PhysRevLett.114.243002}
  {\bibfield  {journal} {\bibinfo  {journal} {Phys. Rev. Lett.}\ }\textbf
  {\bibinfo {volume} {114}},\ \bibinfo {pages} {243002} (\bibinfo {year}
  {2015})}\BibitemShut {NoStop}%
\bibitem [{\citenamefont {Glaetzle}\ \emph {et~al.}(2015)\citenamefont
  {Glaetzle}, \citenamefont {Dalmonte}, \citenamefont {Nath}, \citenamefont
  {Gross}, \citenamefont {Bloch},\ and\ \citenamefont {Zoller}}]{Glaetzle2015}%
  \BibitemOpen
  \bibfield  {author} {\bibinfo {author} {\bibfnamefont {A.~W.}\ \bibnamefont
  {Glaetzle}}, \bibinfo {author} {\bibfnamefont {M.}~\bibnamefont {Dalmonte}},
  \bibinfo {author} {\bibfnamefont {R.}~\bibnamefont {Nath}}, \bibinfo {author}
  {\bibfnamefont {C.}~\bibnamefont {Gross}}, \bibinfo {author} {\bibfnamefont
  {I.}~\bibnamefont {Bloch}}, \ and\ \bibinfo {author} {\bibfnamefont
  {P.}~\bibnamefont {Zoller}},\ }\href@noop {} {\bibfield  {journal} {\bibinfo
  {journal} {Phys. Rev. Lett.}\ }\textbf {\bibinfo {volume} {114}},\ \bibinfo
  {pages} {173002} (\bibinfo {year} {2015})}\BibitemShut {NoStop}%
\bibitem [{\citenamefont {Henkel}\ \emph {et~al.}(2010)\citenamefont {Henkel},
  \citenamefont {Nath},\ and\ \citenamefont {Pohl}}]{Henkel2010}%
  \BibitemOpen
  \bibfield  {author} {\bibinfo {author} {\bibfnamefont {N.}~\bibnamefont
  {Henkel}}, \bibinfo {author} {\bibfnamefont {R.}~\bibnamefont {Nath}}, \ and\
  \bibinfo {author} {\bibfnamefont {T.}~\bibnamefont {Pohl}},\ }\href {\doibase
  10.1103/PhysRevLett.104.195302} {\bibfield  {journal} {\bibinfo  {journal}
  {Phys. Rev. Lett.}\ }\textbf {\bibinfo {volume} {104}},\ \bibinfo {pages}
  {195302} (\bibinfo {year} {2010})}\BibitemShut {NoStop}%
\bibitem [{\citenamefont {Pupillo}\ \emph {et~al.}(2010)\citenamefont
  {Pupillo}, \citenamefont {Micheli}, \citenamefont {Boninsegni}, \citenamefont
  {Lesanovsky},\ and\ \citenamefont {Zoller}}]{Pupillo2010}%
  \BibitemOpen
  \bibfield  {author} {\bibinfo {author} {\bibfnamefont {G.}~\bibnamefont
  {Pupillo}}, \bibinfo {author} {\bibfnamefont {A.}~\bibnamefont {Micheli}},
  \bibinfo {author} {\bibfnamefont {M.}~\bibnamefont {Boninsegni}}, \bibinfo
  {author} {\bibfnamefont {I.}~\bibnamefont {Lesanovsky}}, \ and\ \bibinfo
  {author} {\bibfnamefont {P.}~\bibnamefont {Zoller}},\ }\href {\doibase
  10.1103/PhysRevLett.104.223002} {\bibfield  {journal} {\bibinfo  {journal}
  {Phys. Rev. Lett.}\ }\textbf {\bibinfo {volume} {104}},\ \bibinfo {pages}
  {223002} (\bibinfo {year} {2010})}\BibitemShut {NoStop}%
\bibitem [{\citenamefont {He}\ and\ \citenamefont {Hofstetter}(2011)}]{He2011}%
  \BibitemOpen
  \bibfield  {author} {\bibinfo {author} {\bibfnamefont {L.}~\bibnamefont
  {He}}\ and\ \bibinfo {author} {\bibfnamefont {W.}~\bibnamefont
  {Hofstetter}},\ }\href {\doibase 10.1103/PhysRevA.83.053629} {\bibfield
  {journal} {\bibinfo  {journal} {Phys. Rev. A}\ }\textbf {\bibinfo {volume}
  {83}},\ \bibinfo {pages} {053629} (\bibinfo {year} {2011})}\BibitemShut
  {NoStop}%
\bibitem [{\citenamefont {Gei{\ss}ler}\ \emph {et~al.}(2015)\citenamefont
  {Gei{\ss}ler}, \citenamefont {Vasic},\ and\ \citenamefont
  {Hofstetter}}]{Geissler2015}%
  \BibitemOpen
  \bibfield  {author} {\bibinfo {author} {\bibfnamefont {A.}~\bibnamefont
  {Gei{\ss}ler}}, \bibinfo {author} {\bibfnamefont {I.}~\bibnamefont {Vasic}},
  \ and\ \bibinfo {author} {\bibfnamefont {W.}~\bibnamefont {Hofstetter}},\
  }\href@noop {} {\bibfield  {journal} {\bibinfo  {journal}
  {arXiv:1509.06292v1}\ } (\bibinfo {year} {2015})}\BibitemShut {NoStop}%
\bibitem [{\citenamefont {Santos}\ \emph {et~al.}(2000)\citenamefont {Santos},
  \citenamefont {Shlyapnikov}, \citenamefont {Zoller},\ and\ \citenamefont
  {Lewenstein}}]{Santos2000}%
  \BibitemOpen
  \bibfield  {author} {\bibinfo {author} {\bibfnamefont {L.}~\bibnamefont
  {Santos}}, \bibinfo {author} {\bibfnamefont {G.~V.}\ \bibnamefont
  {Shlyapnikov}}, \bibinfo {author} {\bibfnamefont {P.}~\bibnamefont {Zoller}},
  \ and\ \bibinfo {author} {\bibfnamefont {M.}~\bibnamefont {Lewenstein}},\
  }\href {\doibase 10.1103/PhysRevLett.85.1791} {\bibfield  {journal} {\bibinfo
   {journal} {Phys. Rev. Lett.}\ }\textbf {\bibinfo {volume} {85}},\ \bibinfo
  {pages} {1791} (\bibinfo {year} {2000})}\BibitemShut {NoStop}%
\bibitem [{\citenamefont {Honer}\ \emph {et~al.}(2010)\citenamefont {Honer},
  \citenamefont {Weimer}, \citenamefont {Pfau},\ and\ \citenamefont
  {B\"{u}chler}}]{Honer2010}%
  \BibitemOpen
  \bibfield  {author} {\bibinfo {author} {\bibfnamefont {J.}~\bibnamefont
  {Honer}}, \bibinfo {author} {\bibfnamefont {H.}~\bibnamefont {Weimer}},
  \bibinfo {author} {\bibfnamefont {T.}~\bibnamefont {Pfau}}, \ and\ \bibinfo
  {author} {\bibfnamefont {H.~P.}\ \bibnamefont {B\"{u}chler}},\ }\href
  {\doibase 10.1103/PhysRevLett.105.160404} {\bibfield  {journal} {\bibinfo
  {journal} {Phys. Rev. Lett.}\ }\textbf {\bibinfo {volume} {105}},\ \bibinfo
  {pages} {160404} (\bibinfo {year} {2010})}\BibitemShut {NoStop}%
\bibitem [{\citenamefont {Johnson}\ and\ \citenamefont
  {Rolston}(2010)}]{Johnson2010}%
  \BibitemOpen
  \bibfield  {author} {\bibinfo {author} {\bibfnamefont {J.~E.}\ \bibnamefont
  {Johnson}}\ and\ \bibinfo {author} {\bibfnamefont {S.~L.}\ \bibnamefont
  {Rolston}},\ }\href {\doibase 10.1103/PhysRevA.82.033412} {\bibfield
  {journal} {\bibinfo  {journal} {Phys. Rev. A}\ }\textbf {\bibinfo {volume}
  {82}},\ \bibinfo {pages} {033412} (\bibinfo {year} {2010})}\BibitemShut
  {NoStop}%
\bibitem [{\citenamefont {Maucher}\ \emph {et~al.}(2011)\citenamefont
  {Maucher}, \citenamefont {Henkel}, \citenamefont {Saffman}, \citenamefont
  {Kr\'olikowski}, \citenamefont {Skupin},\ and\ \citenamefont
  {Pohl}}]{Maucher2011}%
  \BibitemOpen
  \bibfield  {author} {\bibinfo {author} {\bibfnamefont {F.}~\bibnamefont
  {Maucher}}, \bibinfo {author} {\bibfnamefont {N.}~\bibnamefont {Henkel}},
  \bibinfo {author} {\bibfnamefont {M.}~\bibnamefont {Saffman}}, \bibinfo
  {author} {\bibfnamefont {W.}~\bibnamefont {Kr\'olikowski}}, \bibinfo {author}
  {\bibfnamefont {S.}~\bibnamefont {Skupin}}, \ and\ \bibinfo {author}
  {\bibfnamefont {T.}~\bibnamefont {Pohl}},\ }\href {\doibase
  10.1103/PhysRevLett.106.170401} {\bibfield  {journal} {\bibinfo  {journal}
  {Phys. Rev. Lett.}\ }\textbf {\bibinfo {volume} {106}},\ \bibinfo {pages}
  {170401} (\bibinfo {year} {2011})}\BibitemShut {NoStop}%
\bibitem [{\citenamefont {Li}\ \emph {et~al.}(2012)\citenamefont {Li},
  \citenamefont {Hamadeh},\ and\ \citenamefont {Lesanovsky}}]{Hamadeh2012}%
  \BibitemOpen
  \bibfield  {author} {\bibinfo {author} {\bibfnamefont {W.}~\bibnamefont
  {Li}}, \bibinfo {author} {\bibfnamefont {L.}~\bibnamefont {Hamadeh}}, \ and\
  \bibinfo {author} {\bibfnamefont {I.}~\bibnamefont {Lesanovsky}},\ }\href
  {\doibase 10.1103/PhysRevA.85.053615} {\bibfield  {journal} {\bibinfo
  {journal} {Phys. Rev. A}\ }\textbf {\bibinfo {volume} {85}},\ \bibinfo
  {pages} {053615} (\bibinfo {year} {2012})}\BibitemShut {NoStop}%
\bibitem [{\citenamefont {Henkel}\ \emph {et~al.}(2012)\citenamefont {Henkel},
  \citenamefont {Cinti}, \citenamefont {Jain}, \citenamefont {Pupillo},\ and\
  \citenamefont {Pohl}}]{Henkel2012}%
  \BibitemOpen
  \bibfield  {author} {\bibinfo {author} {\bibfnamefont {N.}~\bibnamefont
  {Henkel}}, \bibinfo {author} {\bibfnamefont {F.}~\bibnamefont {Cinti}},
  \bibinfo {author} {\bibfnamefont {P.}~\bibnamefont {Jain}}, \bibinfo {author}
  {\bibfnamefont {G.}~\bibnamefont {Pupillo}}, \ and\ \bibinfo {author}
  {\bibfnamefont {T.}~\bibnamefont {Pohl}},\ }\href {\doibase
  10.1103/PhysRevLett.108.265301} {\bibfield  {journal} {\bibinfo  {journal}
  {Phys. Rev. Lett.}\ }\textbf {\bibinfo {volume} {108}},\ \bibinfo {pages}
  {265301} (\bibinfo {year} {2012})}\BibitemShut {NoStop}%
\bibitem [{\citenamefont {Balewski}\ \emph {et~al.}(2014)\citenamefont
  {Balewski}, \citenamefont {Krupp}, \citenamefont {Gaj}, \citenamefont
  {Hofferberth}, \citenamefont {L\"{o}w},\ and\ \citenamefont
  {Pfau}}]{Balewski2014}%
  \BibitemOpen
  \bibfield  {author} {\bibinfo {author} {\bibfnamefont {J.~B.}\ \bibnamefont
  {Balewski}}, \bibinfo {author} {\bibfnamefont {A.~T.}\ \bibnamefont {Krupp}},
  \bibinfo {author} {\bibfnamefont {A.}~\bibnamefont {Gaj}}, \bibinfo {author}
  {\bibfnamefont {S.}~\bibnamefont {Hofferberth}}, \bibinfo {author}
  {\bibfnamefont {R.}~\bibnamefont {L\"{o}w}}, \ and\ \bibinfo {author}
  {\bibfnamefont {T.}~\bibnamefont {Pfau}},\ }\href {\doibase
  10.1088/1367-2630/16/6/063012} {\bibfield  {journal} {\bibinfo  {journal}
  {New J. Phys.}\ }\textbf {\bibinfo {volume} {16}},\ \bibinfo {pages} {063012}
  (\bibinfo {year} {2014})}\BibitemShut {NoStop}%
\bibitem [{\citenamefont {Carmele}\ \emph {et~al.}(2014)\citenamefont
  {Carmele}, \citenamefont {Vogell}, \citenamefont {Stannigel},\ and\
  \citenamefont {Zoller}}]{Carmele2014}%
  \BibitemOpen
  \bibfield  {author} {\bibinfo {author} {\bibfnamefont {A.}~\bibnamefont
  {Carmele}}, \bibinfo {author} {\bibfnamefont {B.}~\bibnamefont {Vogell}},
  \bibinfo {author} {\bibfnamefont {K.}~\bibnamefont {Stannigel}}, \ and\
  \bibinfo {author} {\bibfnamefont {P.}~\bibnamefont {Zoller}},\ }\href@noop {}
  {\bibfield  {journal} {\bibinfo  {journal} {New J. Phys.}\ }\textbf {\bibinfo
  {volume} {16}},\ \bibinfo {pages} {63042} (\bibinfo {year}
  {2014})}\BibitemShut {NoStop}%
\bibitem [{\citenamefont {Jau}\ \emph {et~al.}(2015)\citenamefont {Jau},
  \citenamefont {Hankin}, \citenamefont {Keating}, \citenamefont {Deutsch},\
  and\ \citenamefont {Biedermann}}]{Jau2015}%
  \BibitemOpen
  \bibfield  {author} {\bibinfo {author} {\bibfnamefont {Y.-Y.}\ \bibnamefont
  {Jau}}, \bibinfo {author} {\bibfnamefont {A.~M.}\ \bibnamefont {Hankin}},
  \bibinfo {author} {\bibfnamefont {T.}~\bibnamefont {Keating}}, \bibinfo
  {author} {\bibfnamefont {I.~H.}\ \bibnamefont {Deutsch}}, \ and\ \bibinfo
  {author} {\bibfnamefont {G.~W.}\ \bibnamefont {Biedermann}},\ }\href
  {http://www.nature.com/doifinder/10.1038/nphys3487} {\bibfield  {journal}
  {\bibinfo  {journal} {Nat. Phys.}\ } (\bibinfo {year} {2015})}\BibitemShut
  {NoStop}%
\bibitem [{\citenamefont {Wang}\ \emph {et~al.}(2014)\citenamefont {Wang},
  \citenamefont {Byrd}, \citenamefont {Simbotin},\ and\ \citenamefont
  {C{\^{o}}t{\'{e}}}}]{Wang2014}%
  \BibitemOpen
  \bibfield  {author} {\bibinfo {author} {\bibfnamefont {J.}~\bibnamefont
  {Wang}}, \bibinfo {author} {\bibfnamefont {J.~N.}\ \bibnamefont {Byrd}},
  \bibinfo {author} {\bibfnamefont {I.}~\bibnamefont {Simbotin}}, \ and\
  \bibinfo {author} {\bibfnamefont {R.}~\bibnamefont {C{\^{o}}t{\'{e}}}},\
  }\href {\doibase 10.1103/PhysRevLett.113.025302} {\bibfield  {journal}
  {\bibinfo  {journal} {Phys. Rev. Lett.}\ }\textbf {\bibinfo {volume} {113}},\
  \bibinfo {pages} {025302} (\bibinfo {year} {2014})}\BibitemShut {NoStop}%
\bibitem [{\citenamefont {W{\"{u}}ster}\ \emph {et~al.}(2011)\citenamefont
  {W{\"{u}}ster}, \citenamefont {Ates}, \citenamefont {Eisfeld},\ and\
  \citenamefont {Rost}}]{Wuster2011a}%
  \BibitemOpen
  \bibfield  {author} {\bibinfo {author} {\bibfnamefont {S.}~\bibnamefont
  {W{\"{u}}ster}}, \bibinfo {author} {\bibfnamefont {C.}~\bibnamefont {Ates}},
  \bibinfo {author} {\bibfnamefont {A.}~\bibnamefont {Eisfeld}}, \ and\
  \bibinfo {author} {\bibfnamefont {J.~M.}\ \bibnamefont {Rost}},\ }\href
  {\doibase 10.1088/1367-2630/13/7/073044} {\bibfield  {journal} {\bibinfo
  {journal} {New J. Phys.}\ }\textbf {\bibinfo {volume} {13}},\ \bibinfo
  {pages} {073044} (\bibinfo {year} {2011})}\BibitemShut {NoStop}%
\bibitem [{\citenamefont {M\"obius}\ \emph {et~al.}(2013)\citenamefont
  {M\"obius}, \citenamefont {Genkin}, \citenamefont {Eisfeld}, \citenamefont
  {W\"uster},\ and\ \citenamefont {Rost}}]{Mobius2013}%
  \BibitemOpen
  \bibfield  {author} {\bibinfo {author} {\bibfnamefont {S.}~\bibnamefont
  {M\"obius}}, \bibinfo {author} {\bibfnamefont {M.}~\bibnamefont {Genkin}},
  \bibinfo {author} {\bibfnamefont {A.}~\bibnamefont {Eisfeld}}, \bibinfo
  {author} {\bibfnamefont {S.}~\bibnamefont {W\"uster}}, \ and\ \bibinfo
  {author} {\bibfnamefont {J.~M.}\ \bibnamefont {Rost}},\ }\href {\doibase
  10.1103/PhysRevA.87.051602} {\bibfield  {journal} {\bibinfo  {journal} {Phys.
  Rev. A}\ }\textbf {\bibinfo {volume} {87}},\ \bibinfo {pages} {051602}
  (\bibinfo {year} {2013})}\BibitemShut {NoStop}%
\bibitem [{\citenamefont {Macr{\`{\i}}}\ and\ \citenamefont
  {Pohl}(2014)}]{Macri2014}%
  \BibitemOpen
  \bibfield  {author} {\bibinfo {author} {\bibfnamefont {T.}~\bibnamefont
  {Macr{\`{\i}}}}\ and\ \bibinfo {author} {\bibfnamefont {T.}~\bibnamefont
  {Pohl}},\ }\href {\doibase 10.1103/PhysRevA.89.011402} {\bibfield  {journal}
  {\bibinfo  {journal} {Phys. Rev. A.}\ }\textbf {\bibinfo {volume} {89}},\
  \bibinfo {pages} {011402(R)} (\bibinfo {year} {2014})}\BibitemShut {NoStop}%
\bibitem [{\citenamefont {G\"{a}rttner}\ \emph {et~al.}(2014)\citenamefont
  {G\"{a}rttner}, \citenamefont {Whitlock}, \citenamefont {Sch\"{o}nleber},\
  and\ \citenamefont {Evers}}]{gaerttner2014}%
  \BibitemOpen
  \bibfield  {author} {\bibinfo {author} {\bibfnamefont {M.}~\bibnamefont
  {G\"{a}rttner}}, \bibinfo {author} {\bibfnamefont {S.}~\bibnamefont
  {Whitlock}}, \bibinfo {author} {\bibfnamefont {D.~W.}\ \bibnamefont
  {Sch\"{o}nleber}}, \ and\ \bibinfo {author} {\bibfnamefont {J.}~\bibnamefont
  {Evers}},\ }\href {\doibase 10.1103/PhysRevA.89.063407} {\bibfield  {journal}
  {\bibinfo  {journal} {Phys. Rev. A}\ }\textbf {\bibinfo {volume} {89}},\
  \bibinfo {pages} {063407} (\bibinfo {year} {2014})}\BibitemShut {NoStop}%
\bibitem [{\citenamefont {Lukin}\ \emph {et~al.}(2001)\citenamefont {Lukin},
  \citenamefont {Fleischhauer}, \citenamefont {Cote}, \citenamefont {Duan},
  \citenamefont {Jaksch}, \citenamefont {Cirac},\ and\ \citenamefont
  {Zoller}}]{Lukin2001}%
  \BibitemOpen
  \bibfield  {author} {\bibinfo {author} {\bibfnamefont {M.~D.}\ \bibnamefont
  {Lukin}}, \bibinfo {author} {\bibfnamefont {M.}~\bibnamefont {Fleischhauer}},
  \bibinfo {author} {\bibfnamefont {R.}~\bibnamefont {Cote}}, \bibinfo {author}
  {\bibfnamefont {L.~M.}\ \bibnamefont {Duan}}, \bibinfo {author}
  {\bibfnamefont {D.}~\bibnamefont {Jaksch}}, \bibinfo {author} {\bibfnamefont
  {J.~I.}\ \bibnamefont {Cirac}}, \ and\ \bibinfo {author} {\bibfnamefont
  {P.}~\bibnamefont {Zoller}},\ }\href {\doibase 10.1103/PhysRevLett.87.037901}
  {\bibfield  {journal} {\bibinfo  {journal} {Phys. Rev. Lett.}\ }\textbf
  {\bibinfo {volume} {87}},\ \bibinfo {pages} {037901} (\bibinfo {year}
  {2001})}\BibitemShut {NoStop}%
\bibitem [{\citenamefont {Schempp}\ \emph {et~al.}(2015)\citenamefont
  {Schempp}, \citenamefont {G\"unter}, \citenamefont {W\"uster}, \citenamefont
  {Weidem\"uller},\ and\ \citenamefont {Whitlock}}]{Schempp2015}%
  \BibitemOpen
  \bibfield  {author} {\bibinfo {author} {\bibfnamefont {H.}~\bibnamefont
  {Schempp}}, \bibinfo {author} {\bibfnamefont {G.}~\bibnamefont {G\"unter}},
  \bibinfo {author} {\bibfnamefont {S.}~\bibnamefont {W\"uster}}, \bibinfo
  {author} {\bibfnamefont {M.}~\bibnamefont {Weidem\"uller}}, \ and\ \bibinfo
  {author} {\bibfnamefont {S.}~\bibnamefont {Whitlock}},\ }\href {\doibase
  10.1103/PhysRevLett.115.093002} {\bibfield  {journal} {\bibinfo  {journal}
  {Phys. Rev. Lett.}\ }\textbf {\bibinfo {volume} {115}},\ \bibinfo {pages}
  {093002} (\bibinfo {year} {2015})}\BibitemShut {NoStop}%
\bibitem [{\citenamefont {Glaetzle}\ \emph {et~al.}(2012)\citenamefont
  {Glaetzle}, \citenamefont {Nath}, \citenamefont {Zhao}, \citenamefont
  {Pupillo},\ and\ \citenamefont {Zoller}}]{Glaetzle2012}%
  \BibitemOpen
  \bibfield  {author} {\bibinfo {author} {\bibfnamefont {A.~W.}\ \bibnamefont
  {Glaetzle}}, \bibinfo {author} {\bibfnamefont {R.}~\bibnamefont {Nath}},
  \bibinfo {author} {\bibfnamefont {B.}~\bibnamefont {Zhao}}, \bibinfo {author}
  {\bibfnamefont {G.}~\bibnamefont {Pupillo}}, \ and\ \bibinfo {author}
  {\bibfnamefont {P.}~\bibnamefont {Zoller}},\ }\href {\doibase
  10.1103/PhysRevA.86.043403} {\bibfield  {journal} {\bibinfo  {journal} {Phys.
  Rev. A}\ }\textbf {\bibinfo {volume} {86}},\ \bibinfo {pages} {043403}
  (\bibinfo {year} {2012})}\BibitemShut {NoStop}%
\bibitem [{\citenamefont {Zhao}\ \emph {et~al.}(2012)\citenamefont {Zhao},
  \citenamefont {Glaetzle}, \citenamefont {Pupillo},\ and\ \citenamefont
  {Zoller}}]{Zhao2012}%
  \BibitemOpen
  \bibfield  {author} {\bibinfo {author} {\bibfnamefont {B.}~\bibnamefont
  {Zhao}}, \bibinfo {author} {\bibfnamefont {A.~W.}\ \bibnamefont {Glaetzle}},
  \bibinfo {author} {\bibfnamefont {G.}~\bibnamefont {Pupillo}}, \ and\
  \bibinfo {author} {\bibfnamefont {P.}~\bibnamefont {Zoller}},\ }\href
  {\doibase 10.1103/PhysRevLett.108.193007} {\bibfield  {journal} {\bibinfo
  {journal} {Phys. Rev. Lett.}\ }\textbf {\bibinfo {volume} {108}},\ \bibinfo
  {pages} {193007} (\bibinfo {year} {2012})}\BibitemShut {NoStop}%
\bibitem [{\citenamefont {Huber}\ and\ \citenamefont
  {B\"{u}chler}(2012)}]{Huber2012}%
  \BibitemOpen
  \bibfield  {author} {\bibinfo {author} {\bibfnamefont {S.~D.}\ \bibnamefont
  {Huber}}\ and\ \bibinfo {author} {\bibfnamefont {H.~P.}\ \bibnamefont
  {B\"{u}chler}},\ }\href {\doibase 10.1103/PhysRevLett.108.193006} {\bibfield
  {journal} {\bibinfo  {journal} {Phys. Rev. Lett.}\ }\textbf {\bibinfo
  {volume} {108}},\ \bibinfo {pages} {193006} (\bibinfo {year}
  {2012})}\BibitemShut {NoStop}%
\bibitem [{\citenamefont {Fleischhauer}\ and\ \citenamefont
  {Marangos}(2005)}]{Fleischhauer2005}%
  \BibitemOpen
  \bibfield  {author} {\bibinfo {author} {\bibfnamefont {M.}~\bibnamefont
  {Fleischhauer}}\ and\ \bibinfo {author} {\bibfnamefont {J.~P.}\ \bibnamefont
  {Marangos}},\ }\href {\doibase 10.1103/RevModPhys.77.633} {\bibfield
  {journal} {\bibinfo  {journal} {Rev. Mod. Phys.}\ }\textbf {\bibinfo {volume}
  {77}},\ \bibinfo {pages} {633} (\bibinfo {year} {2005})}\BibitemShut
  {NoStop}%
\bibitem [{\citenamefont {Xu}\ and\ \citenamefont {DeMarco}(2015)}]{Xu2015}%
  \BibitemOpen
  \bibfield  {author} {\bibinfo {author} {\bibfnamefont {W.}~\bibnamefont
  {Xu}}\ and\ \bibinfo {author} {\bibfnamefont {B.}~\bibnamefont {DeMarco}},\
  }\href@noop {} {\bibfield  {journal} {\bibinfo  {journal} {arXiv:1510.00368}\
  } (\bibinfo {year} {2015})}\BibitemShut {NoStop}%
\bibitem [{\citenamefont {Hensler}\ \emph {et~al.}(2003)\citenamefont
  {Hensler}, \citenamefont {Werner}, \citenamefont {Griesmaier}, \citenamefont
  {Schmidt}, \citenamefont {G{\"{o}}rlitz}, \citenamefont {Pfau}, \citenamefont
  {Giovanazzi},\ and\ \citenamefont {Rzazewski}}]{Hensler2003}%
  \BibitemOpen
  \bibfield  {author} {\bibinfo {author} {\bibfnamefont {S.}~\bibnamefont
  {Hensler}}, \bibinfo {author} {\bibfnamefont {J.}~\bibnamefont {Werner}},
  \bibinfo {author} {\bibfnamefont {A.}~\bibnamefont {Griesmaier}}, \bibinfo
  {author} {\bibfnamefont {P.}~\bibnamefont {Schmidt}}, \bibinfo {author}
  {\bibfnamefont {A.}~\bibnamefont {G{\"{o}}rlitz}}, \bibinfo {author}
  {\bibfnamefont {T.}~\bibnamefont {Pfau}}, \bibinfo {author} {\bibfnamefont
  {S.}~\bibnamefont {Giovanazzi}}, \ and\ \bibinfo {author} {\bibfnamefont
  {K.}~\bibnamefont {Rzazewski}},\ }\href {\doibase 10.1007/s00340-003-1334-0}
  {\bibfield  {journal} {\bibinfo  {journal} {App. Phys. B.}\ }\textbf
  {\bibinfo {volume} {77}},\ \bibinfo {pages} {765} (\bibinfo {year}
  {2003})}\BibitemShut {NoStop}%
\bibitem [{\citenamefont {Carusotto}\ and\ \citenamefont
  {Ciuti}(2013)}]{Carusotto2013}%
  \BibitemOpen
  \bibfield  {author} {\bibinfo {author} {\bibfnamefont {I.}~\bibnamefont
  {Carusotto}}\ and\ \bibinfo {author} {\bibfnamefont {C.}~\bibnamefont
  {Ciuti}},\ }\href {\doibase 10.1103/RevModPhys.85.299} {\bibfield  {journal}
  {\bibinfo  {journal} {Rev. Mod. Phys.}\ }\textbf {\bibinfo {volume} {85}},\
  \bibinfo {pages} {299} (\bibinfo {year} {2013})}\BibitemShut {NoStop}%
\end{thebibliography}%

\end{document}